# Photoacoustic Image Formation Based on Sparse Regularization of Minimum Variance Beamformer

**Roya Paridar,**[1] **Moein Mozaffarzadeh,**[1,2] **Mohammad Mehrmohammadi**[3] **and Mahdi Orooji**[1,*]

[1]*Department of Biomedical Engineering, Tarbiat Modares University, Tehran, Iran*
[2]*Research Center for Biomedical Technologies and Robotics (RCBTR), Institute for Advanced Medical Technologies (IAMT), Tehran, Iran.*
[3]*Department of Biomedical Engineering, Wayne State University, Detroit, MI, USA*
[*]*morooji@modares.ac.ir*

**Abstract:** Delay-and-Sum (DAS) is the most common algorithm used in photoacoustic (PA) image formation. However, this algorithm results in a reconstructed image with a wide mainlobe and high level of sidelobes. Minimum variance (MV), as an adaptive beamformer, overcomes these limitations and improves the image resolution and contrast. In this paper, a novel algorithm, named modified-sparse-MV (MS-MV) is proposed in which a $\ell_1$-norm constraint is added to the MV minimization problem after some modifications, in order to suppress the sidelobes more efficiently, compared to MV. The added constraint can be interpreted as the sparsity of the output of the MV beamformed signals. Since the final minimization problem is convex, it can be solved efficiently using a simple iterative algorithm. The numerical results show that the proposed method, MS-MV beamformer, improves the signal-to-noise ($SNR$) about 19.48 *dB*, in average, compared to MV. Also, the experimental results, using a wire-target phantom, show that MS-MV leads to $SNR$ improvement of about 2.64 *dB* in comparison with the MV.

**OCIS codes:** (110.5120) Photoacoustic imaging; (110.3000) Image quality assessment; (110.2990) Image formation theory; (110.3010) Image reconstruction techniques.

## 1. Introduction

Photoacoustic imaging (PAI) is a non-invasive hybrid medical imaging modality that combines the physics of ultrasound (US) and optical imaging modalities [1, 2]. It uses short pulses to illuminate the imaging medium and consequently, generate the acoustic waves. The major benefit of PAI, in comparison with optical imaging, is its higher penetration depth while a high contrast is achieved. It has several applications such as functional and structural imaging [3], ocular imaging [4], tumor detection [5] and imaging the whole body of small animal [6]. PAI is divided into two categories: Photoacoustic microscopy (PAM) and Photoacoustic tomography (PAT) which is suitable for *in vivo* applications [7]. In PAT, an array sensor which is formed in linear, circular or arc shape, is used to detect the propagated photoacoustic (PA) waves from the surface of the medium [8]. Therefore, a reconstruction algorithm is necessary to construct the absorption distribution of the medium [9]. Due to the high similarity between the US and PAI, beamforming algorithms used in US can be applied to the PAI with some modifications [10]. Delay-and-Sum (DAS) is the most common algorithm used to reconstruct the images in US and PAI due to its low computational complexity. However, the reconstructed images obtained from this algorithm suffers from wide mainlobe and high level of sidelobes. Delay-Multiply-and-Sum (DMAS) is proposed to enhance the image quality compared to DAS [11]. In order to improve the reconstructed image in the term of contrast, in comparison with DMAS, some other algorithms have been proposed [12–14]. Adaptive beamformers, which have a wide range of applications in Radar and Sonar, weight the received PA waves proportional to the characteristics of the signals. Therefore, adaptive beamformers would overcome the limitations exist in non-adaptive beamformers and improve the image quality. One of the adaptive beamformers that has been improved in medical image reconstruction, is minimum variance (MV) algorithm [15], in which the weights are calculated in a way that the interference-plus noise power of the output is minimized while the unit gain at the focal point is retained [16]. The reconstructed images obtained from this algorithm is significantly improved in terms of mainlobe width and sidelobe levels compared to non-adaptive beamformers. Some modifications have been applied to the MV beamformer to improve the image quality. Coherence factor (CF) has been applied to MV to improve the resolution and contrast [17]. Two modifications have been applied to CF in order to suppress the sidelobes and improve the resolution more efficiently compared to the conventional CF [18, 19]. A novel beamforming algorithm, called MV-Based DMAS (MVB-DMAS) is introduced in which the MV beamformer is applied inside the expansion of the non-adaptive DMAS beamformer [20, 21]. Temporal averaging has been employed to improve the contrast while the resolution is retained [22]. Also, Eigenspace-Based Minimum Variance (EIBMV) and its combination with DMAS have been employed in order to improve the image quality [23–25].

Minimum Dispersion Distortionless Response (MDDR) is another beamforming method in which the dispersion term is interpreted as a generalized form of the variance. This algorithm is suitable for non-Gaussian signals and noise, where the necessary information is obtained from higher-order and lower-order statistics of the output. In MDDR minimization problem, the weight is estimated in a way that the expectation of the $\ell_p$-norm ($p \geq 1$) of the output is minimized, while the unit gain at the focal point is retained unity [26]. It is obvious that MDDR would be reduced to MV beamformer for $p = 2$, where the second-order statistics of the output is exploited. Higher order statistics can be exploited for $p \geq 2$ and lower order statistics can be exploited for $p \leq 2$. Generally, the value of $p$ is chosen depending on the characteristics of the noise added to the signals [27, 28].

In order to suppress the sidelobes more efficiently, several beamforming methods have been developed in Radar and Sonar applications by adding a new constraint to the existing optimization problem; In [29], a sparse beamforming method, named $\ell_1$-regularized minimum absolute distortionless response ($\ell_1$-MADR), is introduced in which $\ell_1$-regularization term is added to

MDDR beamformer with the assumption of $p = 1$. It has been shown that this robust algorithm suppresses the sidelobe level more efficiently compared to MV. In [30], a new constraint is added to the MV criterion in order to maximize the Mainlobe-to-Sidelobe Power Ratio (MSPR). Sparse capon (SC) beamforming is another method in which a $\ell_p$-regularization term ($p \leq 1$) is added to the MV beamforming constraint in order to force the sparsity to the beampattern and improve the reconstructed images in terms of sidelobes compared to the pure MV beamformer [31, 32]. In this paper, it is proposed to add a new $\ell_1$-norm sparse constraint to the existing MV beamformer after some modifications in order to enhance the image quality and further suppress the sidelobes. The new added constraint can be interpreted as the sparsity of the output which is forced to the entire beampattern. The optimum weight would be obtained from a simple iterative algorithm used in [29]. The results show that the proposed algorithm, named modified-sparse-MV (MS-MV), leads to a significant image improvement in the term of sidelobe levels compared to MV. The rest of the paper is organized as follows. In section 2, MV beamforming method is described. The proposed method is presented in section 3. The results obtained from the numerical data are presented in section 4. Also, the experimental results are evaluated in section 5. Finally, the discussion and the conclusion are reported in section 6 and 7, respectively.

## 2. Background

### 2.1. Minimum Variance Beamformer

In MV beamforming method, the output of the beamformed signal, $y(k)$, is defined as below:

$$y(k) = \sum_{m=1}^{M} w_m(k) x_m(k - \Delta_m(k)), \tag{1}$$

where $M$ is the number of elements in the linear array sensor, $x_m(k)$ is the generated PA signal received by $m^{th}$ element, $k$ is the time index, $\Delta_m$ is the time delay which is calculated proportional to the distance between the point target and $m^{th}$ element and applied to the received signals, and $w_m(k)$ is the calculated weight corresponding to the delayed received signals. Equation (1) can be written as below:

$$y(k) = W^H(k) X(k), \tag{2}$$

where $W(k) = [w_1(k), w_2(k), \cdots, w_m(k)]^T$ is the array of weights and $X(k) = [x_1(k), x_2(k), \cdots, x_M(k)]^T$ is the array of delayed received signals. The weight is calculated in a way that the signal-to-interference-plus-noise ratio ($SINR$) is maximized [33]:

$$\text{SINR} = \frac{\sigma_s^2 |W(k)^H a|^2}{W(k)^H R(k) W(k)}, \tag{3}$$

where $\sigma_s^2$ is the signal power and $R(k)$ is the spatial covariance matrix of the interference-plus-noise. The maximization of the $SINR$ would be achieved by minimizing the interference-plus-noise power of the output (denominator of (3)), while the unit gain is retained at the focal point:

$$\min_{W(k)} W(k)^H R(k) W(k), \quad \text{s.t.} \quad W(k)^H a = 1, \tag{4}$$

where $a$ is the steering vector and $(.)^H$ denotes the conjugate transpose operator. It should be noted that $a$ is considered as a vector of ones since the received signals are delayed. The optimum weight would be obtained using Lagrange multiplier method:

$$W(k) = \frac{R(k)^{-1} a}{a^H R(k)^{-1} a}. \tag{5}$$

The optimum weight is obtained in a way that the large sidelobes are allowed in directions where the received energy is low. Therefore, the off-axis signals would be suppressed [16]. Note that due to the non-stationary property of the signals in medical US and PA, obtaining the exact spatial covariance matrix is not possible. Therefore, the spatial covariance matrix should be estimated:

$$\hat{R} = \frac{1}{N} \sum_{k=1}^{N} X(k)X(k)^H, \qquad (6)$$

where $N$ is the number of received samples. In medical US and PAI, where obtaining $N$ samples is not possible, the array sensor is divided into $M - L + 1$ overlapping subarrays with length $L$, and are replaced by $N$ received samples in (6). Then, $R(k)$ is estimated by averaging the calculated covariance matrixes of each subarray. This technique, named spatial smoothing, leads to a good estimation of the covariance matrix. The estimated covariance matrix is written as below:

$$\hat{R}(k) = \frac{1}{(M - L + 1)} \sum_{l=1}^{M-L+1} \bar{X}_l(k)\bar{X}_l(k)^H, \qquad (7)$$

where $\bar{X}_l(k) = [x_l(k), x_{l+1}(k), \cdots, x_{l+L-1}(k)]^T$ is the delayed received signals of $l^{th}$ subarray. Note that the smaller $L$ increases the number of subarrays, and consequently a more robust estimation would be achieved. However, the resolution would be decreased due to the reduction of employed signals in each subarray. Therefore, it can be concluded that there is a trade off between the subarray length and the resolution. Also, the diagonal loading ($DL$) can be applied to the spatial covariance matrix to obtain a more robust estimation. To apply $DL$, $\Delta.trace\{R\}$ should be added to the diagonal of the estimated covariance matrix before the weight is calculated. $\Delta$ is a constant parameter that corresponds to the subarray length and is much smaller than $\frac{1}{L}$ [34]. In addition to the spatial smoothing, temporal averaging also can be used in spatial covariance matrix estimation to achieve a better resolution, while the contrast is retained [22]:

$$\hat{R}(k) = \frac{1}{(2K + 1)(M - L + 1)} \sum_{n=-K}^{K} \sum_{l=1}^{M-L+1} \bar{X}_l(k+n)\bar{X}_l(k+n)^H, \qquad (8)$$

where the temporal averaging is used over $2K + 1$ samples. The output beamformed signal would be obtained after the weight is estimated using $\hat{R}(k)$ instead of $R(k)$ in (5) from the following equation:

$$\tilde{y}(k) = \frac{1}{M - L + 1} \sum_{l=1}^{M-L+1} W(k)^H \bar{X}_l(k). \qquad (9)$$

## 3. Proposed method

### 3.1. Sparse MV beamforming

In SC beamforming method, which has a wide range of applications in the field of Radar and Sonar [31], a sparse constraint is applied to the MV minimization problem in order to force the sparsity to the hole beampattern and suppress the sidelobes more efficiently compared to the MV beamformer. The optimization problem is defined as follows:

$$\min_{W} W^H RW + \alpha ||C^H W||_p^p, \quad \text{s.t.} \quad a^H W = 1, \qquad (10)$$

where $\alpha$ is the regularization parameter, $C$ is a matrix which covers $K$ steering vectors sampled from $-90°$ to $90°$. Note that in PAI, the steering vector is considered as a vector of ones in all directions, as mentioned before; therefore, the $C$ would be only a matrix of ones with a dimension of $L \times K$. $||.||_p$ represents the $\ell_p$-norm of a vector. It should be noted that the $\ell_p$-norm of a vector can be represents as follows:

$$||\mathbf{h}||_p = \left(\sum_{n=1}^{N} |h(n)|^p\right)^{\frac{1}{p}}, \tag{11}$$

where $\mathbf{h} = [h(1), h(2), \cdots, h(N)]^T$ is a complex vector. Considering $p = 1$, this minimization problem is convex, and therefore, it can be solved efficiently using CVX MATLAB toolbox [35]. In this paper, it is suggested to add a new $\ell_1$-norm sparse constraint to this minimization problem and generate sparse-MV (S-MV) beamformer in order to enhance the image quality and suppress the sidelobes more efficiently. The new optimization problem is defined as follows:

$$\min_{W_S} W_S^H R W_S + \left(\alpha ||C^H W_S||_1 + \beta ||X^H W_S||_1\right), \quad \text{s.t.} \quad a^H W_S = 1. \tag{12}$$

The parameter $\beta$ is considered as a weighting factor which determines the balance between the MV constraint and the new added sparse constraint. In the other words, the sparsity of the output of the MV beamformed signal is forced to the beampattern by adding the new $\ell_1$-norm constraint. This new added constraint leads to significantly image improvement in terms of sidelobe suppression. Combining the two $\ell_1$-norm constraints, (12) can be expressed as follows:

$$\min_{W_S} W_S^H R W_S + ||B W_S||_1, \quad \text{s.t.} \quad a^H W_S = 1, \tag{13}$$

where $B = [\beta X, \alpha C]^H$. In this work, the optimum weight is obtained using the iterative algorithm used in [29]. In the following, the problem solving procedure is explained.

### 3.1.1. Problem solving

The constraint $||BW_S||_1$ in (13) is considered as $||BW_S||_p^p$ in which $p = 1$. Therefore, it can be rewritten as below:

$$||BW_S||_p^p = \sum_{n=1}^{N} |BW_S(n)|^p = $$
$$\sum_{n=1}^{N} |BW_S(n)|^{p-2} |BW_S(n)|^2 = ||\Phi_1 BW_S||^2, \tag{14}$$

where $\Phi_1$ is a diagonal weighting matrix:

$$\Phi_1 = \text{diag}\left\{|BW_S(1)|^{\frac{p-2}{2}}, \cdots, |BW_S(N)|^{\frac{p-2}{2}}\right\}. \tag{15}$$

The motivation behind this scheme, is to convert the $\ell_p$-norm form into the $\ell_2$-norm form, and therefore make the optimization problem solvable [26]. Referring to (13) and (14), the optimization problem would be expressed as follows:

$$\min_{W_S} W_S^H R W_S + ||\Phi_1 BW_S||^2, \quad \text{s.t.} \quad a^H W_S = 1. \tag{16}$$

From this new form of the minimization problem, the optimum weight would be obtained using the Lagrange multiplier method; First, the objective function $f(W_S, \gamma)$ is created as bellow:

$$f(W_S, \gamma) = W_S^H R W_S + ||\Phi_1 B W_S||^2 - \gamma(a^H W_S - 1) = $$
$$W_S^H R W_S + W_S^H B^H D_1(W_S) B W_S - \gamma(a^H W_S - 1), \tag{17}$$

where

$$D_1(W_S) = \Phi_1^H \Phi_1 = \\ \text{diag}\left\{|BW_S(1)|^{p-2}, \cdots, |BW_S(N)|^{p-2}\right\}. \tag{18}$$

By taking the derivation of the objective function with respect to the $W_S$ and $\gamma$, separately, we have:

$$\frac{\partial f}{\partial W_S} = 0 \Rightarrow RW_S + B^H D_1(W_S)BW_S - \gamma a = 0$$
$$\Rightarrow W_S = \gamma(R + B^H D_1(W_S)B)^{-1}a \tag{19}$$

$$\frac{\partial f}{\partial \gamma} = 0 \Rightarrow a^T W_S = 1 \tag{20}$$

Multiplying two sides of (19) to $(a^T)$, we have:

$$a^T W_S = \gamma a^H (R + B^H D_1(W_S)B)^{-1} a$$
$$\Rightarrow \gamma = \frac{1}{a^H (R + B^H D_1(W_S)B)^{-1} a}, \tag{21}$$

and finally, the optimum weight would be obtained using (21) and (19):

$$W_S = \frac{(R + B^H D_1(W_S)B)^{-1} a}{a^H (R + B^H D_1(W_S)B)^{-1} a}. \tag{22}$$

It should be noted that $D_1$ is weight-dependent, and therefore (22) could not be a closed-form solution. The optimum weight would be achieved using the following iterative algorithm:

$$W_S^{k+1} = \frac{(R + B^H D_1(W_S^k)B)^{-1} a}{a^H (R + B^H D_1(W_S^k)B)^{-1} a}, \tag{23}$$

where $(.)^k$ indicates $k^{th}(k = 0, 1, \cdots)$ step of the iteration procedure. Since this iterative algorithm is not sensitive to the initial weight, the MV beamformer or the Bartlett beamformer can be used to obtain the initial weight. The results show that the optimum weight obtained from (23) leads to image enhancement in terms of sidelobe levels and noise suppression compared to MV.

### 3.2. MS-MV beamforming method

The proposed S-MV minimization problem, consists of two sparse constraints; the first sparse constraint which comes from the SC minimization problem, and the new added constraint which forces the sparsity of the output to the entire beampattern. Analysing the first constraint, $||C^H W_S||_1$, we have concluded that this constraint does not have any effect on the sparsity of the beampattern; the steering vector, which is considered as a vector of ones in PA and US, forces the added constraint to be a constant parameter. In other words, a constant parameter is added to the spatial covariance matrix for all time indexes using this constraint, which can not make any changes to the calculated weight compared to MV (see Appendix). In fact, the improvement of the reconstructed images obtained from S-MV, shown in (12), is mainly due to the second new added constraint. As a result, the first constraint could be omitted while the quality of the reconstructed image remains unchanged. The proposed minimization problem, called modified-sparse-MV (MS-MV), finally would be achieved as follows:

$$\min_{W_{MS}} W_{MS}^H R W_{MS} + \beta ||X^H W_{MS}||_1, \quad \text{s.t.} \quad a^H W_{MS} = 1. \tag{24}$$

The optimum weight ($W_{MS}$) would be achieved similar to the previous minimization problem and the final proposed iterative algorithm would be expressed as below:

$$W_{MS}^{k+1} = \frac{(R + \beta X D_2(W_{MS}^k) X^H)^{-1} a}{a^H (R + \beta X D_2(W_{MS}^k) X^H)^{-1} a}, \tag{25}$$

where $D_2(W_{MS}) = \Phi_2^H \Phi_2$, and

$$\Phi_2 = \text{diag}\left\{|X^H W_{MS}(1)|^{\frac{p-2}{2}}, \cdots, |X^H W_{MS}(N)|^{\frac{p-2}{2}}\right\}. \tag{26}$$

The results in the next section show that the proposed MS-MV beamforming method would suppress the sidelobes more efficiently compared to MV beamformer.

## 4. Numerical Result

### 4.1. Imaging setup

The k-wave MATLAB toolbox is used to simulate the array sensor and the absorbers [36]. An imaging region is designed with the vertical and lateral axis of 70 *mm* and 20 *mm*, respectively. Ten 0.1 *mm* radius spherical absorbers, as initial pressures, are centered on the lateral axis. The absorbers are located along the vertical axis. 5 *mm* distance is considered between each two absorbers, starting from 20 *mm* of the array sensor. A linear array sensor including 128 elements is used to detect the propagated PA signals with the central frequency of 5 *MHZ* and 77% bandwidth. The speed of sound is assumed to be 1540 *m/s*. In all simulation cases, the length of the subarrays is considered $L = M/2$. *DL* is applied to the spatial covariance matrix with the assumption of $\delta = 1/100L$. Also, temporal averaging over 5 samples ($K = 2$) is applied. The number of iteration, $N_{iter}$, is considered 10 for calculating the weights in DS-MV. In order to make the signals similar to the real condition, Gaussian noise is added to the received signals. Finally, Hilbert transform, normalization and log-compression procedure are performed after applying the reconstruction algorithm.

### 4.2. Effect of varying $\alpha$ and $\beta$

Before the qualitative evaluation of the reconstructed images using different algorithms, the effect of varying $\alpha$ and $\beta$ needs to be discussed in order to fix them in the procedure of the simulations. Note that $\alpha$ is a parameter that determines the balance between the first added constraint $||C^H W||_1$ and MV constraint, as mentioned before. It can be concluded that different values of $\alpha$ would not make any changes to the reconstructed images due to the effectiveness of this added constraint to S-MV minimization problem. Therefore, only the different values of $\beta$ needs to be investigated. Consider the reconstructed images of a single point target located at the depth of 250 $mm$ using MS-MV for three different values of $\beta$ ($\geq 0$). From the lateral variations shown in Fig. 1, it can be seen that the lowest level of sidelobes would be obtained using $\beta = 1$. Note that the smaller value of $\beta$ causes the algorithm to behave like MV. In this work, the simulation is performed with the consideration of $\beta = 1$. Due to the effectiveness of the first added constraint in S-MV, and therefore, the equality of S-MV and MS-MV, the reconstruction procedure is performed using MS-MV in the next part of this section.

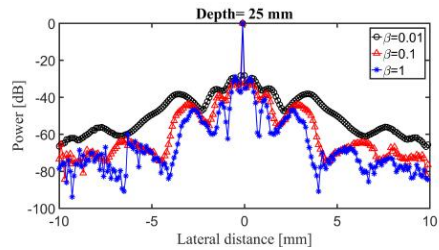

Fig. 1: The lateral variations of a single point target using MS-MV with different values of $\beta$.

### 4.3. Qualitative Evaluation

Figure 2 shows the reconstructed images of the simulated point targets. The reconstructed images using DAS, MV and DS-MV are depicted in Fig. 2(a), Fig. 2(b) and Fig. 2(c), respectively. Considering the images, it is obvious that the non-adaptive DAS beamformer results in the reconstructed image with the lowest quality. MV enhances the image resolution and contrast compared to DAS. Comparing Fig. 2(b) and Fig. 2(c), it can be seen that the proposed DS-MV algorithm improves the image contrast compared to MV, as expected. To have a better evaluation, the lateral variations of the reconstructed images are presented in Fig. 3, at four different depths of imaging. It is demonstrated that the proposed DS-MV leads to image improvement in terms of sidelobe suppression more effectively compared to MV and the non-adaptive DAS beamformers, at all the depths of imaging.

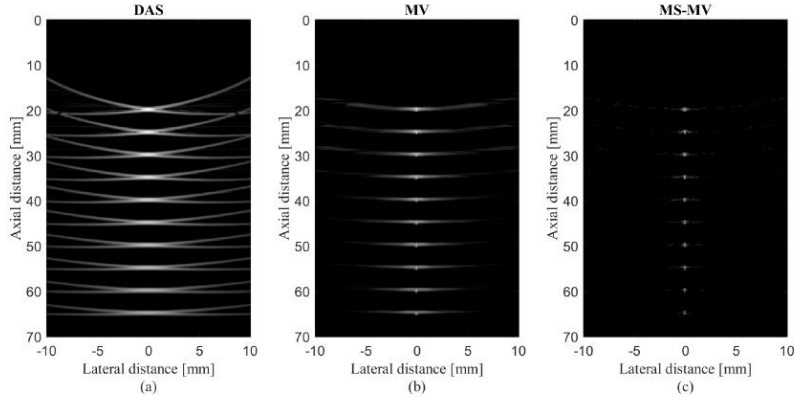

Fig. 2: The numerical reconstructed PA images of 10 point targets using (a) DAS, (b) MV, (c) DS-MV. Noise is added to the received signals having a $SNR$ of 50 $dB$.

### 4.4. Quantitative Evaluation

One of the most common metrics to evaluate the reconstructed images quantitatively, is signal-to-noise ratio ($SNR$), which is calculated as below:

$$SNR = 20 \log_{10} P_{signal}/P_{noise}, \tag{27}$$

where $P_{signal}$ is the difference between the maximum and the minimum intensity of the reconstructed image, and $P_{noise}$ is the standard deviation of the reconstructed image [12, 19]. The $SNR$s of the reconstructed images shown in Fig. 2 are calculated and presented in Table 1. As shown, the DAS leads to the lowest value of $SNR$ due to the existence of high level of sidelobes, as expected. Comparing the calculated $SNR$s corresponding to MV and MS-MV, it can be concluded that the MS-MV beamformer significantly improves the $SNR$ due to the added sparse constraint to the MV criterion. Full-width-half-maximum ($FWHM$) is another metrics to evaluate the performance of the beamformers. This metric is used in order to estimate the spatial resolution of the reconstructed images. The calculated $FWHM$ is presented in Table 2 in all depths of positioning point targets. As can be seen, the highest value of calculated $FWHM$ belongs to DAS. Therefore, it can be concluded that DAS leads to a reconstructed image with the lowest resolution, compared to other beamformers, as expected. Also, It can be concluded that no significant resolution improvement occurred using the proposed MS-MV in comparison with MV since the calculated $FWHM$s are almost similar. It worth to mention that the calculated $FWHM$s are almost constant in all depths of positioning points using MV and MS-MV, as shown in Table 2; the difference between the calculated $FWHM$s at the first and the last depth of positioning points is 2.8 *mm*, 0.04 *mm* and 0.01 *mm* using DAS, MV and MS-MV, respectively. As a result, it can be concluded that the resolution remains almost unchanged using MV and MS-MV in all depths of imaging compared to the non-adaptive DAS beamformer.

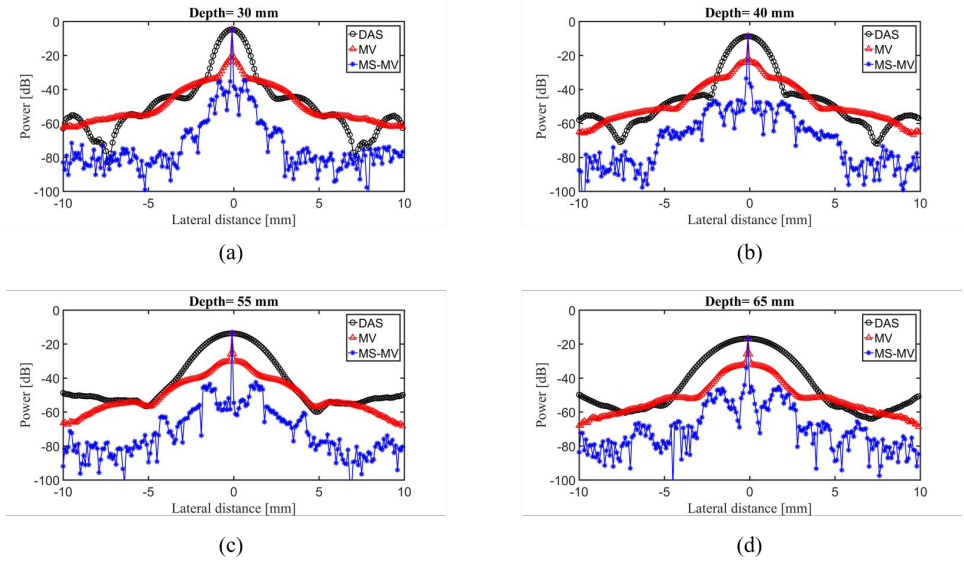

Fig. 3: The lateral variations of the images shown in Fig. 2 at the depth of (a) 30 *mm*, (b) 40 *mm*, (c) 55 *mm* and (d) 65 *mm*. Noise is added to the received signals having a $SNR$ of 50 *dB*.

### 4.5. Imaging at the Presence of High level of Noise

To evaluate the performance of the beamformers at the presence of high level of noise, a Gaussian noise is added to the received signals, resulting the $SNR$ of 10 *dB* . The reconstructed images are shown in Fig. 4. The reconstructed images using DAS, MV and DS-MV are depicted in Fig. 4(a), Fig. 4(b) and Fig. 4(c), respectively. Also, the lateral variations in two different depths of positioning points are shown in Fig. 5. The results show that the reconstructed image using DAS is more affected by the presence of noise. Also, it can be concluded that the reconstructed image using DS-MV is significantly improved in terms of noise reduction compared to other beamformers.

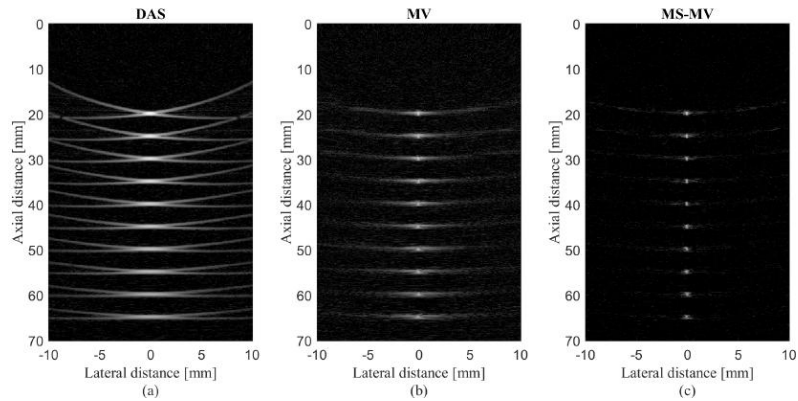

Fig. 4: The reconstructed PA images of 10 point targets using (a) DAS, (b) MV, (c) DS-MV. Noise is added to the received signals having a $SNR$ of 10 $dB$.

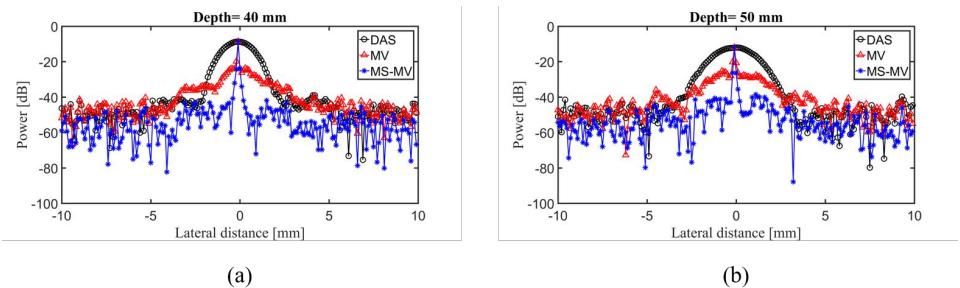

Fig. 5: The lateral variations of the images shown in Fig. 4 at the depth of (a) 40 $mm$ and (b) 50 $mm$. Noise is added to the received signals having a $SNR$ of 10 $dB$.

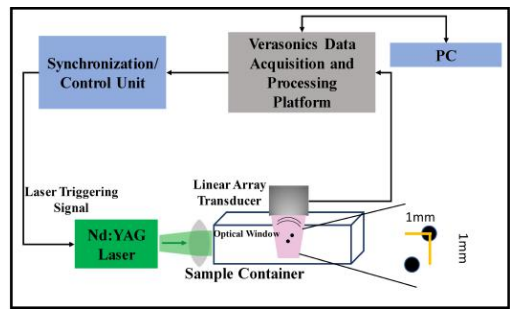

Fig. 6: The schematic of the setup used for the experimental PAI.

## 5. Experimental Results

### 5.1. Wire-target Phantom

To further evaluate the introduced algorithm and its effect on PA image improvement, phantom experiments were performed in which a phantom consists of 1 and 2 light absorbing wires with diameter of 150 *μm* were placed 1 *mm* apart from each other in a water tank. The schematic of the experimental setup is shown in Fig. 6. In the wire-target experiments, we utilized a Nd:YAG pulsed laser, with the pulse repetition rate of 30 $Hz$ at wavelengths of 532 $nm$. A programmable digital ultrasound scanner (Verasonics Vantage 128), equipped with a linear array transducer (L11-4v) operating at frequency range between 4 to 9 $MHz$ was utilized to acquire the PA RF data. A high speed FPGA was used to synchronize the light excitation and PA signal acquisition.

Table 1: The calculated $SNR$ ($dB$) in different depths.

| Depth ($mm$) | DAS | MV | MS-MV |
|---|---|---|---|
| 20 | 35.74 | 53.38 | 79.74 |
| 25 | 33.46 | 52.56 | 77.38 |
| 30 | 31.18 | 52.48 | 75.90 |
| 35 | 29.27 | 50.66 | 72.37 |
| 40 | 27.67 | 50.46 | 68.42 |
| 45 | 26.24 | 48.89 | 67.15 |
| 50 | 24.87 | 48.55 | 66.00 |
| 55 | 23.63 | 48.24 | 63.78 |
| 60 | 22.65 | 47.91 | 62.90 |
| 65 | 21.78 | 47.87 | 62.25 |

Table 2: The calculated $FWHM$ ($mm$) in different depths.

| Depth ($mm$) | DAS | MV | MS-MV |
|---|---|---|---|
| 20 | 0.88 | 0.11 | 0.10 |
| 25 | 1.14 | 0.11 | 0.10 |
| 30 | 1.41 | 0.11 | 0.10 |
| 35 | 1.71 | 0.11 | 0.10 |
| 40 | 1.96 | 0.12 | 0.10 |
| 45 | 2.28 | 0.13 | 0.10 |
| 50 | 2.65 | 0.13 | 0.10 |
| 55 | 2.99 | 0.13 | 0.10 |
| 60 | 3.43 | 0.13 | 0.10 |
| 65 | 3.68 | 0.15 | 0.11 |

5.1.1. Qualitative and Quantitative Evaluation

First, a single wire target was used as the phantom of imaging. The reconstructed images using the concerned beamformers are shown in Fig. 7 where the higher resolution of the MV-based algorithms are clear, compared to the DAS. As demonstrated in Fig. 7(c), a reduced background noise is obtained, in comparison with the image shown in Fig. 7(b), indicating the higher SNR of the MS-MV. To quantitatively evaluate the experimental results, SNR was used where DAS, MV and MS-MV leads to 39.2 $dB$, 40.5 $dB$, 43 $dB$, respectively, which shows the superiority of the proposed method. A more complex target (using two wires as the target) was utilized, and the results are presented in Fig. 8. As demonstrated, DAS results in a low resolution image along with high sidelobes. MV improves the resolution, but the sidelobes still affect the reconstructed image. On the other hand, MS-MV leads to a high resolution while the sidelobes are degraded, and the presence of noise is suppressed in the image, as can be seen in Fig. 8(c). To evaluate the images in detail, lateral variations are presented in Fig. 9 where MS-MV leads to a better resolution compared to DAS while the sidelobes are degraded compared to MV (see the arrows and circles). It can be seen that at the depth of 22 $mm$, using MS-MV, the sidelobes are degraded about 8 $dB$, compared to MV.

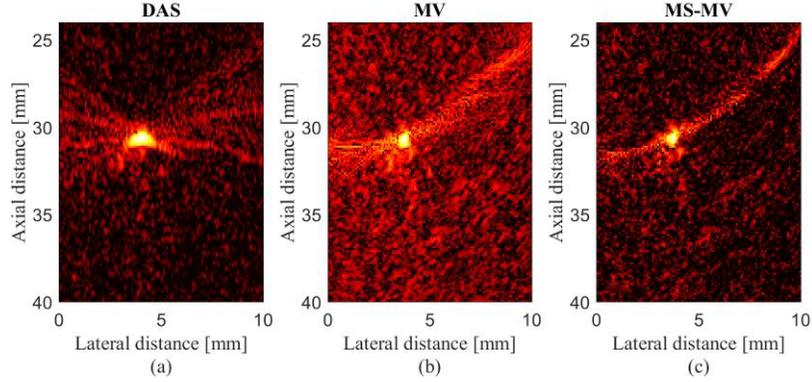

Fig. 7: Reconstructed experimental PA images using (a) DAS, (b) MV and (c) MS-MV ($\beta = 1$). All the images are shown with a dynamic range of 50 dB. A wire target was used as the imaging target.

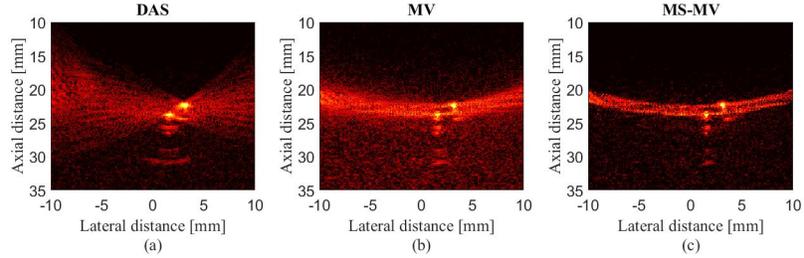

Fig. 8: Reconstructed experimental PA images using (a) DAS, (b) MV and (c) MS-MV ($\beta = 1$). All the images are shown with a dynamic range of 50 dB. Two wires were used as the imaging target.

### 5.2. Ex Vivo Imaging

In this study, an *ex vivo* experimental tissue study have been designed to evaluate the performance of the proposed algorithm. A piece of a chicken breast tissue (about 4 *cm*× 4 *cm*× 3 *cm*) is extracted. Two pencil leads with a diameter of 0.5 *mm* are embedded inside the breast tissue, having an axial distance of about 5 *mm*. Fig 10 shows the photographs of the imaged tissue. The PA signals are collected with a combined linear US/PA imaging probe. The reconstructed images are presented in Fig. 11. As demonstrated, DAS leads to a low resolution while MV provides a higher resolution. On the other hand, the background of the *ex vivo* image obtained by MS-MV is darker which indicates a reduced background noise, in comparison with MV, which indicates the higher SNR of the proposed method. It should be noticed that the sidelobes are lower using MS-MV while the resolution of MV is retained. The lateral variations shown in Fig. 12 clearly proves the lower sidelobes of MS-MV (15 *dB*), in comparison with MV.

### 6. Discussion

The major benefits of the proposed DS-MV algorithm, is the improvement of the reconstructed images in terms of sidelobe levels and noise reduction compared to DAS and MV beamformers. DAS is a non-adaptive algorithm in which the weights are predefined, and therefore, it can not reject the of-axis signals contribution as well as adaptive beamformers. As a results, the reconstructed images using DAS leads to a low quality reconstructed image in terms of resolution and contrast. In the adaptive MV beamformer, the calculated weights are changed proportional to the characteristics of the received signals. Therefore, the reconstructed images would be improved in terms of mainlobe width and sidelobe levels. By adding the new $\ell_1$-norm constraint to the MV criterion, the sidelobes are suppressed more efficiently compared to MV. Consider Fig. 2 and Fig. 3, for instance, where the reconstructed images of the wire phantom and the corresponding lateral variations are shown, respectively. It can be seen that the sparse added constraint leads to a better image quality and more noise reduction compared to MV. Furthermore, the proposed algorithm results in the reconstructed image with a better quality at the presence of high level of noise, as shown in Fig. 4, where 10 *dB* noise is added to the received signals. In fact, the proposed constraint added to the minimization problem in DS-MV, $||X^H W||_1$, forces the sparsity of the output to the whole beampattern; it takes the smaller values to the signals with large amplitudes. It should be noted that most of the large amplitudes exist in mainlobe, but the mainlobe is not affected by this added sparse constraint since the sparse constraint is combined with the other constraint, s.t. $a^H W = 1$. It means that the sparse constraint is performed on the beampattern, while the focal point remains unity. Referring to (11), it is worth to mention that an accurate metric of the sparsity is $\ell_0$-norm, where the non-zero elements of the vector is counted. However, $\ell_0$-norm constraint leads to a non-convex optimization problem. Therefore, $\ell_1$-norm is used to make the minimization problem convex and solvable. Note that in US and PA, the steering vector is considered as a vector of ones in all direction, as mentioned before. That's why the constraint $||C^H W||_1$ does not make any changes to the reconstructed images compared to MV (see APPENDIX for more information). In contrast to US and PA, in Radar and Sonar applications the steering vector varies in each direction, and therefore, this will cause the mentioned constraint to be considered as an efficient constraint which results in sidelobe suppression more efficiently compared to MV. In the other words, in US and PA, there is not any difference between the proposed S-MV and MS-MV due to the effectiveness of the first added constraint, $||C^H W||_1$, exists in S-MV beamformer. However, these two algorithms would behave differently in Radar and Sonar applications. Finally, it should be noted that the computational complexity of DAS, MV and the proposed DS-MV is $O(M)$ and $O(M^3)$ and $O(N_{iter}M^3)$, respectively. Obviously, DAS algorithm can reconstruct the images faster than other beamformers. Comparing MV and DS-MV, it can be concluded that the computational complexity of the proposed DS-MV is comparable with MV.

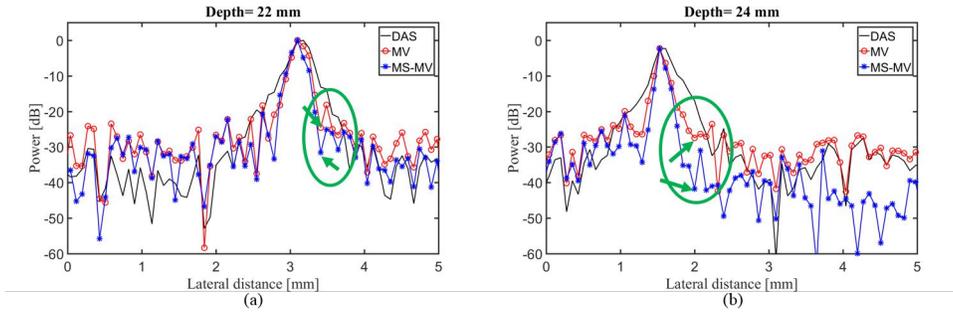

Fig. 9: The lateral variations for the reconstructed experimental PA images shown in Fig. 8. Arrows and circles demonstrate the improvement caused by MS-MV algorithm.

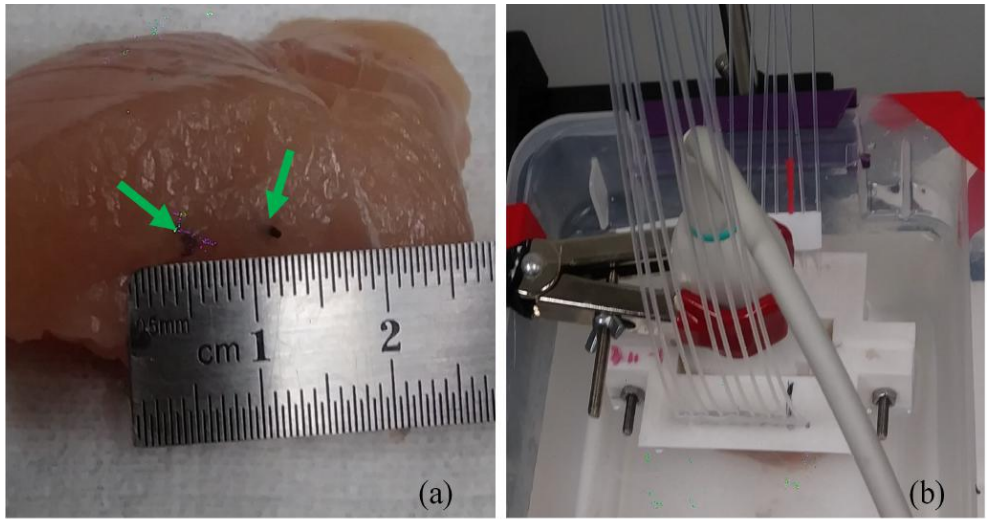

Fig. 10: (a) The phantom used for the experiment. (b) The *ex vivo* imaging setup.

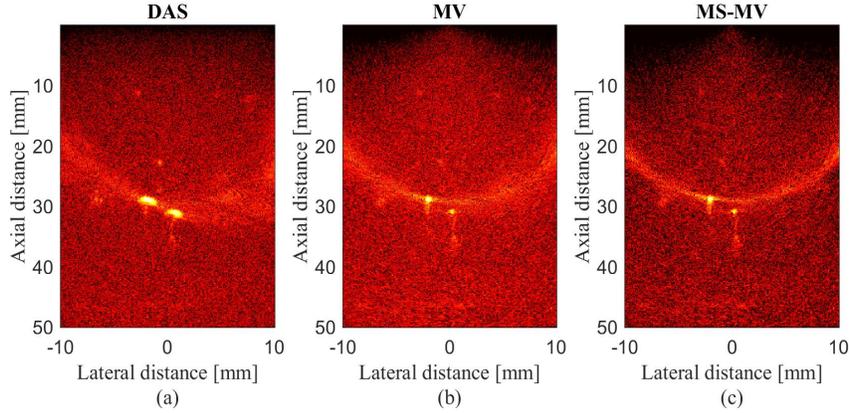

Fig. 11: Reconstructed *ex vivo* images using (a) DAS, (b) MV and (c) MS-MV. A linear-array and the phantom shown in Fig. 10 were used for the experimental design. All the images are shown with a dynamic range of 50 *dB*.

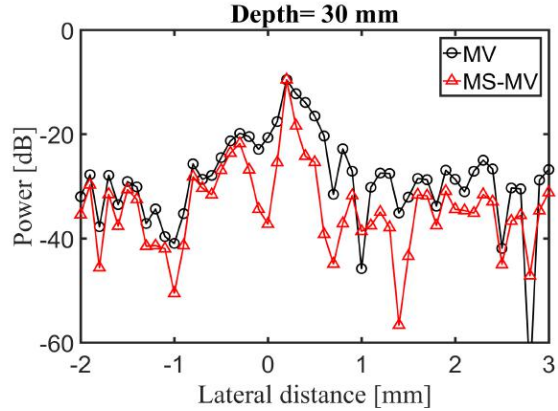

Fig. 12: Lateral variations of the reconstructed images shown in Fig. 11 at the depths of 31.3 *mm*.

## 7. Conclusion

One of the most common algorithms used in PAI, is the non-adaptive DAS beamformer due to its simple implementation. However, the quality of the reconstructed images obtained from this algorithm suffers from the wide mainlobe and high sidelobes. Adaptive beamformers, such as MV, overcomes these limitation by calculating the weights based on the characteristics of the signals and results in an improved reconstructed images in terms of mainlobe width and sidelobe levels. In this paper, it was proposed to add a new constraint to the MV minimization criterion in order to further suppress the sidelobes. The new $\ell_1$-norm added constraint can be interpreted as the sparsity of the output that is forced to the beampattern. The numerical and experimental results showed that the proposed DS-MV algorithm outperforms MV in terms of sidelobe suppression and noise reduction. The quantitative results obtained from DS-MV showed that the calculated $SNR$ was improved about 19.48 $dB$ and 2.64 $dB$ for the simulated point targets and the designed wire phantom, respectively, compared to MV.

## Funding

This research received no specific grant from any funding agency in the public, commercial, or not-for-profit sectors.

## Disclosures

The authors declare that there are no conflicts of interest related to this article.

## APPENDIX

### 7.1. Analysis the Sparse Constraint ($||C^H W||_1$)

In US and PAI applications, the first sparse constraint in S-MV minimization problem, $||C^H W||_1$, does not have any effect on the reconstructed images compared to MV beamformer; the steering vector is considered as a vector of ones in all directions, as mentioned before. This property makes the added constraint to be a constant parameter which is added to the covariance matrix. It can be seen that adding a constant parameter would not make any changes to the estimated weight compared to the estimated weight obtained from MV. For better understanding, consider SC minimization problem in (10), in which the desired constraint is added to MV minimization problem. The optimum weight is obtained from the same method mentioned before:

$$w = \frac{(R + \alpha C D(w) C^H)^{-1} a}{a^H (R + \alpha C D(W) C^H)^{-1} a}, \tag{28}$$

where $D(W) = \text{diag}\left\{|C^H W(1)|^{-1}, \cdots, |C^H W(N)|^{-1}\right\}$. $D(W)$ can be written as below:

$$D(W) = \begin{bmatrix} \left|\sum_{n=1}^{N} w(n)\right|^{-1} & \cdots & 0 \\ \vdots & \ddots & \vdots \\ 0 & \cdots & \left|\sum_{n=1}^{N} w(n)\right|^{-1} \end{bmatrix}_{K \times K} \tag{29}$$

It should be noted that each elements in diagonal of the matrix in (29), equals to the second constraint combined with the first constraint in SC beamformer, $W^H a = 1$. Therefore, $D(W)$ would be a diagonal matrix of ones with dimensions $K \times K$. By expanding the term added to the covariance matrix shown in (28), we have:

$$\alpha CD(W)C^H = \alpha \times$$
$$\begin{bmatrix} 1 & \cdots & 1 \\ \vdots & \ddots & \vdots \\ 1 & \cdots & 1 \end{bmatrix}_{N \times K} \cdot \begin{bmatrix} 1 & \cdots & 0 \\ \vdots & \ddots & \vdots \\ 0 & \cdots & 1 \end{bmatrix}_{K \times K} \cdot \begin{bmatrix} 1 & \cdots & 1 \\ \vdots & \ddots & \vdots \\ 1 & \cdots & 1 \end{bmatrix}_{K \times N} \quad (30)$$
$$= \begin{bmatrix} \alpha N & \cdots & \alpha N \\ \vdots & \ddots & \vdots \\ \alpha N & \cdots & \alpha N \end{bmatrix}_{N \times N}$$

It is obvious from (30) that the term added to the covariance matrix would be a $N \times N$ matrix where each elements of this matrix is a constant parameter, $\alpha N$. As a result, (28) would be rewritten as below:

$$w = \frac{(R + \alpha N)^{-1} a}{a^H (R + \alpha N)^{-1} a}. \quad (31)$$

in which the constant parameter, $\alpha N$, is added to the covariance matrix. In order to show the equality of the calculated weights obtained from (5) and (31), consider the covariance matrix with dimension $2 \times 2$ ($N = 2$) for simplicity, as below:

$$R = \begin{bmatrix} a & b \\ c & d \end{bmatrix} \quad (32)$$

The inverse of (32) would be as follows:

$$R^{-1} = \frac{1}{\underbrace{ad - bc}_{A}} \begin{bmatrix} d & -b \\ -c & a \end{bmatrix} \quad (33)$$

Referring to (5), the estimated weight obtained from MV is as follows:

$$w_{MV} = \frac{R^{-1} a}{a^H R^{-1} a} = \frac{\frac{1}{A} \begin{bmatrix} d - b \\ -c + a \end{bmatrix}}{\frac{1}{A}(d - b - c + a)} \quad (34)$$
$$= \frac{1}{(d - b - c + a)} \begin{bmatrix} d - b \\ a - c \end{bmatrix}$$

Adding the constant term to the covariance matrix, we have:

$$(R + \alpha N) = \begin{bmatrix} a + \alpha N & b + \alpha N \\ c + \alpha N & d + \alpha N \end{bmatrix} \quad (35)$$

The inverse of (35) would be as below:

$$(R + \alpha N)^{-1} = \frac{1}{\underbrace{ad - bc + \alpha N(a + d - b - c)}_{B}} \times$$
$$\begin{bmatrix} d + \alpha N & -b - \alpha N \\ -c - \alpha N & a + \alpha N \end{bmatrix}$$

Referring to (31), the optimum weight obtained from SC beamformer is estimated as follows:

$$w_{SC} = \frac{(R+\alpha N a)^{-1} a}{a^H (R+\alpha N)^{-1} a} = \frac{\frac{1}{B}\begin{bmatrix} d+\alpha N - b - \alpha N \\ -c - \alpha N + a + \alpha N \end{bmatrix}}{\frac{1}{B}(d-b-c+a)} \qquad (36)$$

$$= \frac{1}{(d-b-c+a)} \begin{bmatrix} d-b \\ a-c \end{bmatrix}$$

Comparing (34) and (36), it can be seen that the weigh obtained from MV equals to the weight obtained from SC beamformer in which a sparse constraint is added to the existing minimization problem ($w_{MV} = w_{SC}$). In the other words, it can be concluded that the mentioned sparse constraint does not make any changes to the calculated weight compared to MV in US and PA image reconctruction.